\documentclass[preprint,12pt]{elsarticle}

\usepackage{amssymb}

\journal{Physics of Life Reviews}

\begin{document}

\begin{frontmatter}


\author{Manlio De Domenico\corref{cor1}}
\ead{mdedomenico@fbk.eu}
\address{Fondazione Bruno Kessler, Via Sommarive 18, 38123 Povo (TN), Italy\fnref{label3}}

\title{Multilayer Network Modeling of Integrated Biological Systems\\Comment on ``Network Science of Biological Systems at Different Scales: A Review'' by Gosak et al.\tnoteref{label1}}

\end{frontmatter}



Biological systems, from a cell to the human brain, are inherently complex. A powerful representation of such systems, described by an intricate web of relationships across multiple scales, is provided by complex networks. Recently, several studies are highlighting how simple networks -- obtained by aggregating or neglecting temporal or categorical description of biological data -- are not able to account for the richness of information characterizing biological systems. More complex models, namely multilayer networks, are needed to account for interdependencies, often varying across time, of biological interacting units within a cell, a tissue or parts of an organism. 

Gosak et al~\cite{gosak2017review} review the most recent advances in the application of multilayer networks for modeling complex biological systems, from molecular interactions within a cell to neuronal connectivity of the human brain.

\section{Network Science of Biological Systems}

Biology provides a fertile ground for some of the most exciting applications of Network Science. The essential molecular components of a cell are related by functional interdependencies of different nature (e.g., genetic, physical, etc.) at different scales (e.g., genetic, metabolic, etc. ), making network modeling an essential tool for their modeling and analysis.

Complex networks have improved our understanding of life and disease. On the one hand, the function of a human cell is the result of interacting proteins which are responsible for the underlying modular and hierarchical organization into complexes, organelles and signal transduction pathways that build the human interactome~\cite{huttlin2017architecture}. Because of this interdependency, the alteration of single genes can quickly propagate a perturbation to the protein-protein interaction network, causing abnormal functions in tissues and organs that culminate in diseases. Network medicine~\cite{barabasi2010network} is a developing branch of Network Science which provides theoretical and computational tools for analyzing and predicting the potential effects of such perturbations, by linking genome, transcriptome, proteome and metabolome to human diseases, allowing for the study of the human diseasome, the network of disease-disease interactions~\cite{goh2007human}.

On the other hand, understanding the complex mechanisms behind the function of human brain and cognition is a challenge where promising results have been obtained within the framework of network neuroscience~\cite{bassett2017network}. This developing branch of Network Science allows for modeling and analysis of structural interactions and functional interdependencies within the human brain. The alteration of structural or functional connectivity due, for instance, to localized abnormalities in specific areas of the brain or in their functional relationships, can quickly propagate, causing systemic failures that culminate in neurodegenerative diseases~\cite{stam2006small}, neurological~\cite{chavez2010functional} or psychiatric disorders~\cite{lynall2010functional}.

\section{Multilayer Networks: a Framework for Integrated Biological Systems}

Distinct complex networks provide a fair description of isolated networked systems, consisting of static units which are related by a single type of relationships. Nowadays, such systems are named single-layer networks or, less frequently, simplex networks.

However, biological systems exhibit a higher level of complexity, with interdependencies within and across different networks that can also vary over time. Multilayer network models~\cite{kivela2014multilayer,boccaletti2014structure,wang2015evolutionary,de2016physics} provide a powerful representation~\cite{de2013mathematical} of such systems and allow for the integration of multiple types of interactions among biological units of different types, while reducing loss or aggregation of available information. In this framework, each network is encoded into a different layer of the system, while layers can be coupled each other in a complex way, to resemble complex interaction patterns observed in biology.

Gosak et al~\cite{gosak2017review} propose exciting perpsectives for an even more pervasive use of multilayer network modeling in biology. More specifically, they investigate possible applications to intercellular interaction patterns, where multilayer networks built from simultaneous multicellular recordings allow for integrated network descriptions at multicellular and tissue level. However, the scarcity of experimental data is not the only obstacle to our comprehension of the interplay between structure, dynamics and function of cells.

\subsection{Multi-omics}

Despite the efforts~\cite{mostafavi2010fast,bersanelli2016methods}, integrating different sources of genetic data is still challenging. The ultimate goal is to gain novel insights and new knowledge about life and disease from system-level molecular interactions~\cite{sun2014integrated}. To this aim, multiples sources of omics data encode different layers, representing a biological system as a network of networks. This integrated perspective allows for more predictive performances~\cite{de2015structural,kim2016multi,zitnik2017predicting} and has been shown to better characterize the evolution of complex diseases such as cancer~\cite{tordini2016genome}, as well as to better understand the response to genetic and metabolic perturbations in complex organisms like \emph{E. coli}~\cite{klosik2017}.

\subsection{Connectomics}

Another open challenge concerns with the integration of multiple relationships among units of a nervous system. Only recently multilayer network models have been used to map the connectomes of complex organisms at different scales, from nematoda to \emph{Homo Sapiens} (see \cite{de2017multilayer} for a review). In the case of the model organism \emph{C. elegans}, a multilayer network with alternative modes of interaction (synaptic, gap junction, and neuromodulator) between neurons has been introduced~\cite{de2015muxviz,nicosia2015measuring}, allowing for a better understanding of aminergic and peptidergic modulation of behaviour~\cite{bentley2016multilayer}. Anatomical and functional information has been integrated to gain insights about the macro-scale topology of the Macaque monkey~\cite{crofts2016structure} and the human brain~\cite{battiston2017multilayer}. Temporal~\cite{bassett2011dynamic,vidaurre2017brain} and multi-frequency~\cite{de2016mapping,tewarie2016integrating,yu2017selective,guillon2017loss} decompositions of human brain activity, followed by their successive integration into multilayer networks, have been used to improve our understanding of brain function in cognitive tasks and brain diseases, such as Alzheimer's, Parkinson's or Schizophrenia.

\section{Open Challenges}

Despite the success of multilayer network modeling and analysis in systems biology and systems medicine, some methodological challenges are still to be tackled to build consistent, replicable and reproducible~\cite{drummond2009replicability,ioannidis2009repeatability,ioannidis2011improving,goodman2016does} representations of multi-omics, connectomics and intercellular interactions as the ones central to the review of Gosak et al~\cite{gosak2017review}.

In fact, the extraction of functional connectivity patterns, being them of molecular, neuronal or intercellular origin, still suffers from i) the lack of a robust choice of statistical similarity descriptors (e.g., cross-correlation, spectral coherence, information-theoretic measures) and ii) the absence of objective criteria to threshold the resulting similarity matrices for filtering the observed correlations.

Network scientists are investigating alternatives to traditional approaches, grounded on the development of objective null models based, for instance, on random matrix theory~\cite{macmahon2013unbiased} or problem-specific topological principles~\cite{fallani2017topological}. In fact, the lack of objective and robust methodologies -- for dealing with simplex and multilayer networks inferred from measurements -- might alter the interpretation of results based on the calculation of centrality measures and the determination of meso-scale structures~\cite{fallani2014graph}. 

Further research is definitively needed to assess the statistical variability of network descriptors and make the interpretation of results more robust.




\end{document}